# Physics of Caustics and Protein Folding:

# Mathematical Parallels

Walter Simmons

Department of Physics and Astronomy

University of Hawaii at Manoa

Honolulu, HI 96822

Joel L. Weiner

Department of Mathematics

University of Hawaii at Manoa

Honolulu, HI 96822

Abstract

The energy for protein folding arises from multiple sources and is not large in total. In spite of the many specific successes of energy landscape and other approaches, there still seems to be some missing guiding factor that explains how energy from diverse small sources can drive a complex molecule to a unique state. We explore the possibility that the missing factor is in the geometry.

A comparison of folding with other physical phenomena, together with analytic modeling of a molecule, led us to analyze the physics of optical caustic formation and of folding behavior side-by-side. The physics of folding and caustics is ostensibly very different but there are several strong parallels. This comparison emphasizes the mathematical similarity and also identifies differences.

Since the 1970's, the physics of optical caustics has been developed to a very high degree of mathematical sophistication using catastrophe theory. That kind of quantitative application of catastrophe theory has not previously been applied to folding nor have the points of similarity with optics been identified or exploited.

A putative underlying physical link between caustics and folding is a torsion wave of non-constant wave speed, propagating on the dihedral angles $\Phi$ and $\Psi$ found in an analytical model of the molecule.

Regardless of whether we have correctly identified an underlying link, the analogy between caustic formation and folding is strong and the parallels (and differences) in the physics are useful.

Introduction

The translation of genetic information, which is encoded in the DNA, into uniquely folded proteins defines a central mechanism in all living cells.  The first stages of the process, entailing the translation of the information into an amino acid sequence in the protein, have been understood for a long time.  The final step, the folding of the protein into a unique native state, has been an intensely active research field for more than a half century [1-5]. The fact that the energy available for folding arises from multiple sources and is small in total has been a vexatious problem.  Compounding the energy issue are the facts that, in the random coil state, the hydrophobic and hydrophilic forces do not have well defined directions and also the protein competes with the solvent for Hydrogen bond formation.

We surveyed various physical phenomena for similar situations: strong geometrical pattern formation, independence from perturbations, and small energy requirements.  Optical caustic formation [6-7] stood out as an interesting candidate for comparison with folding.

Optical caustics are a very well understood set of physical phenomena [8-9] that require no energy (besides a source of light) and they are stable against perturbations.  The physics of folding and of caustics is ostensibly different but there are interesting similarities as well as some differences.  The physics of caustics is understood with a high level of mathematical sophistication.  A purpose of this paper is to present arguments supporting the idea of applying aspects of the theory of optical caustics to folding.

We examine the parallels and differences between folding and caustics and identify the possible mathematical similarities.  We attempt to reconcile the differences that we identified.

Parallels of Folding and Caustics

As stated, above, one of the most vexatious aspects of the protein folding problem is that there is only a small amount of energy available to explain the rapid spontaneous folding that occurs in water and in cells. In spite of this energy problem, folding shows some remarkable regularities [10,2]. Each protein proceeds to fold to a unique three dimensional final state, which may be constructed from a finite number of sub-structures, and the final state of the folding process is relatively insensitive to modest perturbations in temperature, pH, etc.

A caustic is an envelope of light rays that arises when light shines upon some object that refracts or reflects light from many points. For example, if a glass of water, which is standing on a table, is illuminated by a bright distant source, a caustic appears on the table-top. The caustic has bright lines, arcs, and cusps. The pattern is three-dimensional and appears different according to the distance and orientation of the surface upon which the light falls.

The optical field, including the caustic, is an interference pattern which requires no additional energy to form. There are only a finite number of caustics that can be uniquely characterized geometrically. Also, the formation of caustics is strikingly insensitive to perturbations.

The theory of caustics, which we summarize in a subsequent section, entails the application of mathematics to the propagation of electromagnetic waves subject to various boundary conditions. One of our motivations for comparing caustics and folding is the appearance of waves and solitons in an analytic molecular model, which we describe elsewhere [11]. We shall not make specific use of these waves in this paper.

## Caustics and Singularity Theory

We shall begin with qualitative and modestly quantitative explanation of what a caustic is and what it means for a caustic to be stable by considering caustics on the bed of a stream, such as seen in a photo on Wikipedia. How does it happen that the shapes of the bright spots on the stream bed remain so unchanged with time when the surface of the water that generates the shapes is so complicated and is continuously changing with time?

Before we can continue, the reader needs a sense of what we mean when we talk about singularities. It is a word that characterizes a point, or set of points, at which some qualitative change in a function or family of functions occurs.

Familiar examples of singularities in functions are given by points of thermodynamic potentials in (P, V, T) space where phase transitions occur.

On the other hand, consider a 1-parameter family of real-valued functions on the real line that depends on one parameter. For certain values of the parameter, there may be a maximum and minimum that can be made to approach one another as the parameter approaches a particular value. When the parameter reaches that value, the maximum and minimum coalesce creating a point at which both the first and second derivatives vanish. For subsequent values of the parameter, the maximum and minimum no longer exist. The value of the parameter at which the change just described occurs is a singular point.

In the case of the light striking the stream bed, the time-delay function of light rays arriving at the stream bottom is the function we study with respect to singularity theory.

Two kinds of parameters generally enter such discussions, state parameters and control parameters, the control parameter taking the role of what we referred to as a parameter in the preceding example. Points at which the first and appropriate second derivatives with respect to the state parameters vanish are generally referred to as singular points or "non-Morse" points. In the case caustics on the streambed, the positions of points on the bed are the control parameters and the coordinates of points at the surface of the stream are state parameters.

Catastrophe theory is a very precise theory of the behavior of families of functions at or around singular points. R. Thom and others, who developed the theory, showed that in many common circumstances that this behavior can be described by simple polynomial functions called elementary catastrophes and furthermore demonstrated the fact that small perturbations of the elementary catastrophes do not change the essential features of the catastrophe, i.e., the elementary catastrophes are stable against small perturbations.

Two important areas of application are to fields of extremals from variational problems that depend on parameters and to potential functions that depend on parameters. In the first instance, think of light rays and Fermat's Principle and in the second thermodynamic potentials and potential energy surfaces. In fact, we remark that catastrophe theory has been applied to energy landscape potentials [12,13]. It is with the first case that we will focus our attention.

We turn now to the general theory of optical caustics. The light field is obtained from the phase, or time-delay, function which appears in the standard Fresnel integral. The phase function, $\Phi$, depends upon information about the light source and upon the components of the optical system that reflect and refract the light; these dependences are often included in the state variables. The phase function also depends upon control parameters, which include the coordinates of the image.

There are three types of points in the light field.

(1) Dark points at which destructive interference reduces the intensity to zero.

(2) Morse Singularities. Here $\nabla \Phi = 0$ and the determinant of the second partial derivatives of $\Phi$ is non-zero $\Phi_{j,k} \equiv \partial_j \partial_k \Phi$, where all the derivatives are with respect to state variables.  The light amplitude at such points depends upon the source amplitude, upon a propagation factor and upon a factor of $\left|Det\Phi_{j,k}\right|^{-1/2} \neq 0$, which is evaluated at the observation screen.

These points include images; focused images are very sensitive to perturbations of the refracting/reflecting surface.

(3) Non-Morse Singularities.  Points at which $\nabla\Phi = 0$ and $Det(\Phi_{j,k}) = 0$.  In geometrical optics $\left|Det\Phi_{j,k}\right|^{-1/2} \to 0$ and the intensity is singular.  In wave optics the intensity goes like $(\frac{1}{\lambda})$ to a power that depends upon the details (where $\lambda$ is the wavelength).

The literature has not standardized upon terminology and there are many synonyms for various technical terms.  We shall use the term 'singularity' for non-Morse points.

An important result of catastrophe theory is that, for a sufficiently small number of control parameters, the number of different caustics is fixed by the number of elementary catastrophes. The shapes of the caustics are insensitive to the kinds of small perturbations that would seriously degrade an image.  (We remark that caustics are three dimensional and are seen only in cross sections, which depend upon the viewing arrangement, hence there appear to be many caustic shapes.  Moreover, there may be multiple caustics that overlap.)

If a caustic were built up statistically, one photon at a time as they are detected on the image screen, we could nevertheless define the optical phase as described above.

Folding and Caustics Compared

Denatured protein molecules (random coils) are at equilibrium in a suitable solvent. The solvent is altered and the molecules proceed to fold to a unique native state. We shall focus upon two-stage folding to the secondary structure; the random coil folds directly to the native state.

Very little energy is involved and the native state is not changed by various perturbations during folding. Nevertheless, the molecule moves toward a specific stable conformation. Large changes to the folding environment, (i.e. values of temperature, pressure, etc. that lie outside certain ranges), disrupt the progression to the native state.

The information that defines these native state conformations is contained in the sequence. Sub-segments of biologically active proteins seem to fall into a finite number of domains.

The sensitivity of the native state to changes in the amino acid sequence is highly varied. Many molecules retain their shape in spite of substantial substitutions while others change their space significantly upon even one, well placed substitution. Additionally, recent research has shown that a high degree of symmetry in the sequence in adjacent domains leads to mis-folding.[15]

The detailed motion of the molecule during folding is known through simulations and experiments in some cases. No general theory of the motion is currently established.

Let us compare these observations to caustics.

The caustics form when the light refracts or reflects from a physical object, which we shall simply call a 'lens'. The lens might be a glass of water or a curved reflecting surface.

The optical field of the caustic takes on a three dimensional shape, which is well defined in terms of the properties of the lens. We view the light by inserting a viewing surface (a 'screen') and the pattern seen depends upon the screen and where it is placed.

No energy is involved.

The optical field is precisely defined at the singular points up to a change of coordinates.  If the parameters describing the lens or screen are smoothly changed then the shape of the caustic remains unchanged near the singular point.

Outside of the caustic, the behavior is different; the structural stability of the optical field is lost.  Also, one of the features of caustics is that they can be destroyed by application of symmetry to the lens.

Let us next take up some issues.

Caustics arise from electromagnetic waves.  Protein folding is not a wave and interference based phenomenon (unless some yet undiscovered quantum effect is involved).

The nature of the state parameters, which might constitute the lens, is unknown.  The same can be said for the control parameters.  Also, at first glance, the protein folding problems appears to be of a discrete nature in some respects, but catastrophe theory requires that we deal with smooth functions.

The number of caustics is small compared to the number of observed protein motifs and domains.

If the initial phase of the folding of some part of a molecule is governed by a catastrophe, the later bond-forming phase may disguise the structurally stable form.  (In a much more limited way, the screen does this to the light).

We conclude this section by observing that, in spite of some issues, the caustic-folding analogy has much to recommend it.  Caustics and folding share similar behavior in energy, they form a finite number of stable shapes in noisy environments, they are both sensitive to certain external forces and symmetries but not sensitive to other forces, and there is an information filtering process in which shapes are picked out of many possibilities.

## Addressing the Issues

At the end of the previous section, we stated our main conclusion. In this section we discuss the issues raised. This is necessarily more speculative.

Our research was motivated, in part, by the appearance of torsion waves of variable wave speed moving back and forth in our analytical model of a protein molecule. The model is not sufficiently developed to model caustic-like processes. We shall make no further use of our analytic model here but we shall assume that a mathematical description of folding somehow parallel to optical caustic theory is possible.

We might speculate that torsion waves propagate smoothly on much of the molecule but encounter short segments with heterogeneous amino acid properties that act in some way on the passing wave. These lens like segments create singular points in the wave phase.

Since folding is not simply local, we suggest two approaches to estimating a length to correspond to a lens structure.

(1) Imagine one end of the molecule pinned down. With two degrees of freedom per residue, three residues have six degrees of freedom. This is sufficient freedom (ignoring steric hindrance, etc.) to describe a small neighborhood in which the next residue can have any position and orientation.

(2) The length of a conserved sequence is suggestive of a range. Mutations that take place outside the range do not change the folding. This suggests a maximum range of roughly 10 residues.

Although the ideas discussed here are insufficient to make calculations, we can look at some data to illustrate the idea.

We assume that a large section of the molecule is relatively homogeneous in amino acid polarity and that in a small sub-section the polarity is heterogeneous. Two possible parameters that could describe this sub-section are the position and polarity of a residue. For the considerations that follow, it seems fairly reasonable to regard the density of polarity as a state variable and position as a control variable.

De Alba, Rico, and Jimenez [16] measured various peptides and showed that a sequence of four or five residues directs a hairpin. After weighting the residues with their polarities we find that two or three residues direct the turn. This would correspond to either a cusp or swallowtail catastrophe, both of which have sharp cusps that might correspond to the hairpin.

We conclude from this section that the issues can, in principle, be addressed.

Conclusions

Motivated by the minimal energy in folding, by the insensitivity of the native form to perturbations during folding, and by the appearance of various waves in our analytic models of protein molecules, we examined folding and optical caustic formation side-by-side.

We found that the analogy demonstrates clear parallels in the physics, which we listed at the end of a prior section.  The quantitative theory of caustic formation, using catastrophe theory, is highly developed and available in many text books.  We argue that the theoretical structure of caustic theory can be useful in folding research.

We also found some issues between caustics and folding, and we discussed how some additional assumptions regarding amino acid heterogeneity could make it possible to develop a geometric theory of folding using catastrophe theory by using this parallel with the theory of optical caustics.

One issue is that caustics are a wave phenomenon (although geometric optics also gives a complete picture of caustics).  We speculated that torsion waves on the molecule backbone, disrupted by heterogeneities in the arrangement of amino acids, form singular points which direct the folding into elementary geometric catastrophes in short segments. Bond formation may subsequently alter the shape.

Recently, there have been many advances in technology in the study of protein folding [187]  which create ample opportunities to test the theoretical ideas discussed here.

Should the geometric application of singularity theory, paralleling the development of the theory of optical caustic formation, prove to be successful, an important advantage of such a formulation would be a reduction in the number of free parameters needed to understand folding.

# References

1.) Dill, K.A., Ozkan, S.B, Shell, M.S, and Weikl, T.R, The Protein Folding Problem, AnnRevBioph 37, 289 -316, (2008).

2.) Rose, George D., Fleming, Patrick J., Banavar, Jayanth R., and Maritan, Amos, A backbone-based theory of protein folding PNAS 103 (45), 16623 (2006)

3.) Onuchic,J.N. & Wolynes,P.G. Theory of Protein Folding, Current Opinion in Structural Biology, 14, 70, (2004).

4.) Dagget,V. & Fersht,A.R. Is there a unifying mechanism for protein folding, TRENDS in Biochemical Sciences 28 (1), 18, (2009).

5.) Vendruscolo,M. et al Small-World View of the Amino Acids that Play a Key Role in Protein Folding, PRE 65, 061910 (2002)

6.) Berry,M.V. Beyond Raimbows, Current Science 59, (21 & 22), 1175 (1990).

7.) Berry, M.V., Waves and Thom's Theorem, Advances in Physics, 25, 1, (1976).

8.) Gilmore, R., Catastrophe Theory for Scientists and Engineers, John Wiley & Sons, New York, (1981).

9.) Petters,A.O., Levine,H. & Wambsgnass,J. Singularity Theory and Gravitational Lensing, Birkhauser, Boston, (2001).

10.) K. A. Dill, S. Bromberg, K. Yue, K. M. Fiebig, D. P. Yee, P. D. Thomas, and H. S. Chan, Principles of protein folding—a perspective from simple exact models, Protein Sci. 1995 April; 4 (4): 561–602.

11.) Simmons, W. & Weiner,J., Protein Folding: A New Geometric Analysis, arXiv:0809.2079.

12.) David J. Wales, A Microscopic Basis for the Global Appearance of Energy Landscapes, Science, 293 no. 5537 pp. 2067-2070, (2001).

13.) Wales, D.J. Energy Landscapes, Cambridge University Press, Cambridge, UK 2003.

14.) X. Krokidis et al Characterization of elementary chemical processes by catastrophe theory, J. Chem. Phys. A. (1997), 101, 7277.

15.) Borgia,M.B. et al Single-molecule fluorescence reveals sequence-specific misfolding in multidomain proteins, Nature 474, 662 (2011).

16.) De Alba,E, Rico, M. and Angeles Jimenez,M., The turn sequence direct beta strand alignment in designed beta hairpins, Protein Science 8, 2234 (1999)

17.) Bartlett, A.I. & Radford, S.E. An expanding arsenal of experimental methods yields an explosion of insights into protein folding mechanisms, Nature Structural & Molecular Biology, 16, 582 (2009)